\documentclass[a4paper,11pt]{article}
\pdfoutput=1
\usepackage{graphicx}
\usepackage{xcolor, graphics}
\usepackage{amsmath,amssymb,amsfonts}
\usepackage{acronym}
\usepackage{mathrsfs}
\usepackage{units}
\usepackage{enumerate}
\usepackage{etoolbox}
\usepackage{float,slashed}
\usepackage{enumitem}
\usepackage{mathtools}
\usepackage[normalem]{ulem}
\usepackage{booktabs}
\usepackage{bbm,tabularx}

\def\beq{\begin{equation}\begin{aligned}}
\def\eeq{\end{aligned}\end{equation}}

\def\dd{{\rm d}}

\usepackage{jcappub}
\usepackage[T1]{fontenc}
\usepackage{booktabs,array}
\usepackage{tikz}
\usetikzlibrary{arrows.meta,calc}
\title{ADROIT: Agentic Dark matter Reinterpretation and Operator Inference Toolkit}
\author[a,b]{James Unwin}
\affiliation[a]{Department of Physics \& AIRSHIP Lab \& LQuTE,\\ University of Illinois Chicago, Chicago, IL 60607,~USA}
\affiliation[b]{Rudolf Peierls Centre for Theoretical Physics,\\ University of Oxford, Oxford OX1 3PU, UK}

\abstract{
We present a tool for expediting the study of dark matter models via AI: \textsc{Adroit}. The system builds on established deterministic dark matter analysis tools developed over the last couple of decades. \textsc{Adroit} combines two functions: First, it uses language-model agents to expedite the construction of dark matter models and their comparison with experimental constraints and the wider literature. Second, it combines established dark matter analysis tools with internal calculations to create a unified workflow for analysing these models. \textsc{Adroit} presents four distinct capabilities, aiding the user in: i).~fitting effective interactions to direct detection signals; ii).~constructing particle models from natural-language descriptions; iii).~fitting specified particle models to signals; and iv).~testing models against relic density, collider, direct- and indirect-detection constraints. The system has a preloaded library of common models and experimental constraints, and the capacity for users to define additional models, limits and signals. Language-model agents can assist with model construction, preparation of experimental inputs, literature review, data import, and review and interpretation of the results. The objective is to expedite human-led physics in which researchers define the questions and review the proposed models and inputs, while agents assist their preparation and numerical codes evaluate their physical consequences. Expert review of the physical assumptions and results remains essential.}

\begin{document}
\maketitle

\section{Introduction}

Different microscopic models can present as the same macroscopic observable signal, especially for finite data. In particular, a small number of nuclear recoils can admit several dark matter interpretations.
Conversely, an observational null result can constrain broad families of models.
Interpreting a dark matter signal requires a connection between the observed events, the underlying particle interactions and the constraints from other experiments. We present \textsc{Adroit} \cite{Adroit}, a research assistant that brings these calculations together. Moreover, 
language-model agents extend this analysis by constructing candidate models, searching the literature, importing constraints and reviewing the results, with the researcher approving the proposed inputs and providing oversight.

A range of public tools already performs the individual calculations. Codes such as WIMpy NREFT \cite{WIMpyNREFT}, DMFormFactor \cite{Anand:2013yka} and WimPyDD \cite{Jeong:2021bpl}  evaluate nuclear responses and recoil rates for dark matter-nucleon interactions. DirectDM \cite{Bishara:2017nnn} relates quark-level couplings to these interactions, while DDCalc \cite{GAMBITDarkMatterWorkgroup:2017fax} provides experimental likelihoods for direct detection. MadGraph \cite{Alwall:2011uj,Alwall:2014hca} computes collider production and SModelS \cite{Altakach:2024jwk} reinterprets LHC searches via simplified models. DarkCast \cite{Ilten:2018crw} recasts dark photon searches for other vector bosons.  MadDM \cite{Ambrogi:2018jqj} and micrOMEGAs \cite{Belanger:2026asz}  compute annihilation rates and the thermal relic abundance. Global-fit frameworks such as GAMBIT, through its DarkBit module \cite{GAMBITDarkMatterWorkgroup:2017fax}, combine dark matter observables with other constraints in a statistical analysis. Here our emphasis is on the steps surrounding these calculations, and on the aspiration to use agentic interfaces to provide seamless progress from an idea, through its formulation, to its comparison with data.

\textsc{Adroit}, \underline{A}gentic \underline{D}ark matter
\underline{R}einterpretation and \underline{O}perator
\underline{I}nference \underline{T}oolkit, provides four main
capabilities for analysing freeze-out dark matter:
\begin{itemize}
\item Fitting non-relativistic effective interactions to
direct detection recoil signals.
\item Constructing candidate Lagrangians from natural-language
descriptions.
\item Fitting library and user-defined particle models to
experimental signals.
\item Testing models against relic-abundance, collider, direct-
and indirect detection.
\end{itemize}
These connect the two directions of dark matter phenomenology, inferring a model or constraint from observations and predicting the consequences of a theory.
\textsc{Adroit} combines established public packages \cite{WIMpyNREFT,Anand:2013yka,Bishara:2017nnn,An:2025diamx,Alwall:2011uj,Alwall:2014hca,Altakach:2024jwk,Ilten:2018crw} with internal calculations, and uses independent codes \cite{Jeong:2021bpl,GAMBITDarkMatterWorkgroup:2017fax,Belanger:2026asz,Ambrogi:2018jqj} to cross-check the numerical implementation. In particular, the LZ, XENONnT and PandaX-4T likelihoods are evaluated with DIAMX \cite{An:2025diamx}, which models each experiment's published analysis.
 The principal new ingredient introduced here is the use of language-model agents to streamline and unify analyses, and to extend the capabilities of the base code to user-defined models and experimental inputs.

The toolkit is intended for experienced physicists. Choosing an appropriate (possibly simplified) model, judging the relevance of additional states and deciding when an experimental recast needs a more detailed acceptance calculation remain a core part of the researcher's analysis. The aim of \textsc{Adroit} is to expedite the work of experienced researchers. 
Moreover, by performing numerical calculations with deterministic code rather than language models, the system improves the reproducibility and reliability of these calculations while reducing the computational demands on the language-model agents (the advantage over direct prompting is illustrated  in Appendix~\ref{A}).
 Figure~\ref{fig:agentic-flow} shows how a request passes through the agents and numerical calculations to the final report. We highlight that this approach is part of a growing effort to use agents in theoretical physics \cite{Alexander:2026lpw,Saad:2026pan,Lucente:2026kgh,Elsharkawy:2026dte,Plehn:2026gxv,Xiao:2026gal,Diefenbacher:2026azr,Guo:2026ifi,Singh:2026emc,Wang:2026jjn,Alexander:2026fhe,Agrawal:2026lvg,Morgensztern:2026hij,Niarchos:2026fbt,Faroughy:2026dkj,Hammad:2026ged,Menzo:2026eet,Chandra:2026njg}.


The paper is structured as follows, we outline the workflow of the system in Section~\ref{S2}. In Section~\ref{S3} we discuss how {\sc{Adroit}} applies constraints to a given model, followed by the methods for inferring dark matter interactions from a given recoil signal in Section~\ref{S4}. In Section \ref{S5} we demonstrate the capabilities of {\sc{Adroit}} through worked examples connecting model construction, fits to direct detection signals and tests against experimental constraints. Comparisons with established benchmark calculations provide checks of the numerical implementation (see Appendix \ref{A}). These examples illustrate the practical role of agentic assistance throughout the analysis.  We give our concluding remarks in Section \ref{S6}.

\begin{figure}[p]
\centering
\begin{tikzpicture}[
 font=\small,
 box/.style={draw=black!55,fill=black!2,line width=.55pt,rounded corners=1.5pt,
 align=center,inner xsep=7pt,inner ysep=8pt,text width=12.8cm},
 half/.style={box,text width=5.65cm,minimum height=1.8cm},
 human/.style={box,fill=blue!5,draw=blue!35!black},
 optional/.style={half,fill=black!1,dashed},
 flow/.style={-{Stealth[length=2mm]},draw=black!65,line width=.65pt,rounded corners=2pt},
 optionalflow/.style={flow,dashed},
 note/.style={font=\footnotesize,align=center,fill=white,inner sep=3pt}
]
\node[human] (person) at (0,0)
 {\textbf{Researcher defines the question}\\
 Model or signal; parameters, halo assumptions and comparisons};
\node[optional] (agents) at (-3.4,-2.2)
 {\textbf{Optional AI assistance}\\
 Model proposals and completions;\\source discovery and extraction};
\node[half] (inputs) at (3.4,-2.2)
 {\textbf{Specify and review inputs}\\
 Fields and interactions;\\experimental data and assumptions};
\draw[optionalflow] (person.south) -- (0,-1.05) -| (agents.north);
\draw[flow] (person.south) -- (0,-1.05) -| (inputs.north);
\draw[optionalflow] (agents.east) -- (inputs.west);
\node[half] (theory) at (-3.4,-4.6)
 {\textbf{Constrain a model}\\
 Model checks and parameter scan;\\thermal-relic solutions};
\node[half] (signal) at (3.4,-4.6)
 {\textbf{Fit and interpret a signal}\\
 Fits, spectral classes, significance\\and null-search compatibility};
\draw[flow] (inputs.south) -- (3.4,-3.4) -| (theory.north);
\draw[flow] (inputs.south) -- (signal.north);
\draw[optionalflow] (signal.south) -- (3.4,-5.95) -- (-3.4,-5.95) -- (theory.south);
\node[note] at (0,-5.95) {construct and test a particle model};
\node[box] (physics) at (0,-7.2)
 {\textbf{Shared deterministic calculations}\\
 Dedicated models or generic amplitudes; recoil rates and likelihoods;\\
 abundance evolution and complementary searches};
\draw[flow] (theory.west) -- (-6.95,-4.6) |- ([yshift=3mm]physics.west);
\draw[flow] (signal.east) -- (6.95,-4.6) |- ([yshift=3mm]physics.east);
\node[half,minimum height=2.6cm] (tools) at (-3.4,-10.0)
 {\textbf{Applicable external calculations}\\
 DirectDM: matching\\
 DIAMX: xenon likelihoods\\
 MadGraph: production\\SModelS and DarkCast: recasts};
\node[optional,minimum height=2.6cm] (checks) at (3.4,-10.0)
 {\textbf{Independent cross-checks}\\
 micrOMEGAs, MadDM, DDCalc\\
 within their overlapping domains};
\draw[flow] (tools.north) -- (physics.south -| tools.north);
\draw[optionalflow] (checks.north) -- (physics.south -| checks.north);
\node[box] (claims) at (0,-12.8)
 {\textbf{Combine the physical tests}\\
 Fit, abundance and experimental comparisons; reviewed inputs;\\
 exclusions and unresolved comparisons with their reasons};
\draw[flow] ([yshift=-3mm]physics.west) -- (-7.25,-7.5) |- (claims.west);
\node[box] (saved) at (0,-15.15)
 {\textbf{Saved study and evidence}\\
 Results, assumptions, report and figures with their inputs;\\
 inspect, export or continue the saved calculation};
\draw[flow] (claims.south) -- (saved.north);
\draw[optionalflow] (saved.east) -- (7.4,-15.15) |- (person.east);
\node[note,rotate=90] at (7.75,-7.1) {researcher-directed follow-up};
\end{tikzpicture}
\caption{From a model or recoil signal to a physical interpretation. Solid arrows show the main calculation path, dashed arrows show agent assistance, independent numerical comparisons and further studies requested by the researcher. The abundance solver is internal, with micrOMEGAs and MadDM used for cross-checks, and DDCalc for direct detection. Models and imported experimental inputs pass through review.}
\label{fig:agentic-flow}
\end{figure}
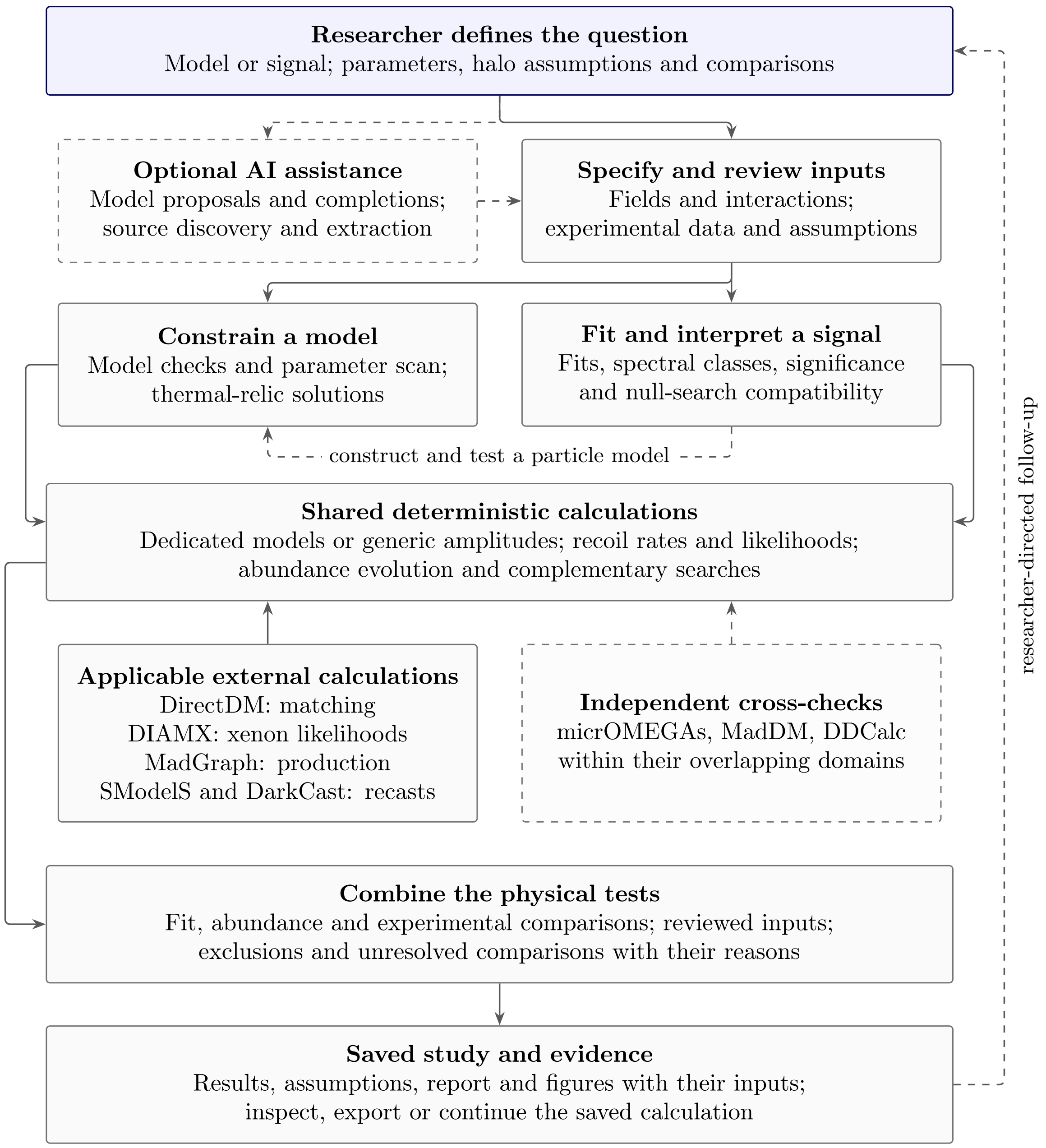

\vspace{-1mm}
\section{Workflow}
\label{S2}
\vspace{-1mm}

Starting from an idea, a Lagrangian, or a publication,  an agent constructs a candidate model. Alternatively the user may choose from a library of common models.  Agents search for relevant constraints, extract published limits and review the completed analysis. The researcher reviews and approves newly constructed models. Imported experimental inputs also require review, either by the researcher or, where the researcher has enabled it, by a second language-model check. The built-in libraries provide a convenient starting point and can be used directly, without invoking a language model, or modified by the user or by an agent.

The model defines the fields, gauge representations, masses and interactions. \textsc{Adroit} checks gauge invariance and hermiticity, diagonalises mass matrices and constructs the interaction vertices. Models in the built-in library use dedicated calculations. 
For models outside the library, \textsc{Adroit} uses the specified interactions to calculate tree-level amplitudes. (The model and
interactions can be exported in UFO format for
use in other physics codes.)
These interactions also determine which experimental limits are applied. Interpreting a mediator search, for example, requires the mediator's production rate, width, and branching fractions. A limit derived for one decay pattern can be applied only where the proposed model predicts the corresponding signal. Four-fermion interactions are checked against their effective-theory cutoff, and coannihilating states are checked for the conversions needed to maintain chemical equilibrium. If a necessary limit or computation is absent from the system or beyond its current capability, the report identifies the calculations still needed to verify a given model.

\section{Constraining a Model}
\label{S3}

One of the primary functions of \textsc{Adroit} is the study of experimental constraints on dark matter models.
The path towards constraints begins by implementing the dark matter scenario within the model library either manually or via an agent.
Once a model has been implemented within the library, the system builds its interaction structure.
On initiating a run, for each interaction and dark matter mass, the code calculates the recoil-energy spectrum using the nuclear-recoil rate (where $\xi_T$ is the mass fraction of isotope $T$) \cite{Anand:2013yka}
\beq
 \frac{\dd R}{\dd E_R}=\frac{\rho_\chi}{m_\chi}
 \sum_T\frac{\xi_T}{m_T}\int_{v>v_{\min}}\dd^3v\,
 f_{\rm lab}(\mathbf v)\,v\,\frac{\dd\sigma_T}{\dd E_R}.
 \label{eq:rate}
\eeq

The present version of \textsc{Adroit} focuses on thermal freeze-out (WIMP) dark matter when evaluating the relic abundance. To identify the parameter space for which the dark matter reproduces the observed dark matter abundance via freeze-out, \textsc{Adroit} averages the annihilation rates over the thermal distribution and solves for the abundance.\footnote{For freeze-out, the system uses the Saikawa-Shirai equation of state \cite{Saikawa:2018rcs} to account for the temperature-dependent energy and entropy densities of the Standard Model plasma, including the QCD crossover. This affects the expansion and thermal history of the early Universe, and hence the dark matter
 relic abundance.}  
The abundance calculation assumes standard thermal freeze-out of a single stable dark matter component, including coannihilating partners where chemical equilibrium is maintained \cite{Griest:1990kh,Edsjo:1997bg,Garny:2017rxs}. Across the scanned parameter space, the code evaluates direct- and indirect-detection, collider, flavour and cosmological constraints \cite{Slatyer:2015jla,Sabti:2019mhn,Planck:2018vyg},  and identifies the points that reproduce the observed dark matter abundance: $\Omega_\chi h^2=0.12$ \cite{Planck:2018vyg}.
The code also interfaces with SModelS~3.2.1, DarkCast and public gamma-ray constraints \cite{Altakach:2024jwk,Ilten:2018crw,McDaniel:2023bju}. Collider production rates are obtained from precomputed tables or, where needed, from MadGraph \cite{Alwall:2011uj,Alwall:2014hca}. Each experimental input is stored as a record containing its source, numerical data, and assumptions. The researcher can add records to a personal library alongside {\sc Adroit}'s standard library.

Astrophysical and thermal assumptions can also affect the resulting viable parameter space. By default, the calculation samples an uncertainty region in the Standard Halo Model (SHM). Specifically, the local density $\rho_0$ is log-normal with median $0.40$ GeV cm$^{-3}$ and $\sigma_{{\rm ln}\rho}=0.35$, the SHM speed parameter is $v_0=233\pm20$ km s$^{-1}$, and the escape velocity is $v_{\rm esc}=528\pm30$ km s$^{-1}$. These are our adopted distributions for exploring halo uncertainty. The central velocities follow \cite{McMillan:2016jtx,Deason:2019tpl}, while the median density is rounded to $0.40$ GeV cm$^{-3}$. The stated widths are our broadened choices rather than the quoted measurement errors. We treat the three parameters as independent and test combinations covering 95\% of their joint probability distribution.
When studying the constraints or signal prospects of a specific model, the user can select either SHM or SHM++, where SHM++ includes the anisotropic Gaia Sausage component \cite{Evans:2018bqy}.  Recoil fits default to each experiment's published halo assumptions, unless a different benchmark is selected. For xenon-based searches we follow the recommended conventions of \cite{Baxter:2021pqo}: $\rho_0=0.3$ GeV cm$^{-3}$, $v_0=238$ km s$^{-1}$ and $v_{\rm esc}=544$ km s$^{-1}$.

The output includes the viable parameter space, a table of evaluated points and a report explaining the exclusions and any unresolved issues. Selecting a point reveals its abundance, the comparison with each search and the assumptions used. Each study records the model, experimental inputs, and calculation settings, which can be exported together with the numerical results.
A point is labelled \emph{excluded} when the included calculations establish a conflict with an observational limit under the adopted assumptions, and \emph{allowed} when it passes the included requirements. 
It remains \emph{pending} when the calculations needed to establish either conclusion are unresolved.
When the halo uncertainty region is sampled, a point is excluded or allowed only if it is so for every sampled halo. Otherwise it remains pending, and the report states whether the outcome varies across the region (in which case it is labelled \emph{halo-sensitive}) or if it has been established only at the central SHM.
Explaining a signal additionally requires a fit to the signal events using parameters in the viable model space, as we discuss below.

\vspace{-2mm}
\section{Explaining a Signal}
\label{S4}
\vspace{-2mm}

\textsc{Adroit} offers two distinct paths towards understanding excesses in direct detection experiments. The first compares a library of interaction hypotheses with the observed events and ranks their fits. The second starts from a specified particle model and determines which values of its physical parameters can reproduce the signal. Both use the recoil calculation of Eq.~(\ref{eq:rate}) and the likelihood function described below.

The first path ``signal best fit'' ingests the details of the signal (e.g.~via a paper upload) and then compares it to a library of candidate interactions. Specifically, \textsc{Adroit} implements a collection of candidate interactions based on the standard non-relativistic operator basis
\cite{Fitzpatrick:2012ix,Anand:2013yka} as well as the relativistic benchmarks considered by LZ \cite{LZ:2026axp}.
For direct detection, the relativistic interactions are matched to non-relativistic operators, with running and matching through DirectDM where applicable \cite{Fan:2010gt,Fitzpatrick:2012ix,Bishara:2017nnn}.
The recoil calculation uses $\mathcal O_1$ and\footnote{Following \cite{Fitzpatrick:2012ix,Anand:2013yka}, we omit $\mathcal O_2$, which does not arise at leading order in the non-relativistic reduction.} $\mathcal O_3$-$\mathcal O_{15}$, using the published nuclear responses \cite{Anand:2013yka,Catena:2015uha},
whose numerical coefficients are taken from WIMpy NREFT
\cite{WIMpyNREFT} and evaluated within \textsc{Adroit}. We use the spin-dependent structure factors of \cite{Klos:2013rwa} where applicable. The recoil spectra are cross-checked against WimPyDD \cite{Jeong:2021bpl}.
For each candidate interaction, \textsc{Adroit} predicts the recoil spectrum that the experiment would measure, accounting for the detector response, efficiency, and exposure. Within the chosen reconstructed-energy window $W$, this prediction is described by the expected signal count $s$ and the distribution of signal events in energy, $f_s(E)$. The latter is normalised so that
\beq
\int_W f_s(E)\dd E=1.
\eeq

The candidate search evaluates a specified grid of dark matter masses and interaction choices. At each grid point, the relative interaction coefficients and other parameters that determine the spectral shape are fixed. The initial fit compares these candidate spectra with the observed events while allowing the overall signal count to vary. Thus, the mass is scanned over the grid, while the normalisation is fitted separately for each candidate. The fit uses both the number of events and their measured energies through an extended likelihood
\beq
 \ln\mathcal L=-(s+\theta B)
 +\sum_{i=1}^{N_{\rm obs}}\ln[s f_s(E_i)+\theta b(E_i)]
 +\ln C(\theta),
 \label{eq:likelihood}
\eeq
where $E_i$ is the reconstructed recoil energy of the $i$-th event. The sum runs over all $N_{\rm obs}$ observed events in the analysis window $W=[E_{\min},E_{\max}]$.
The assumed background spectrum $b(E)$ integrates to an expected background count $B$. The parameter $\theta$ allows this count to vary from its nominal value, with $\theta=1$ corresponding to the assumed background.
When the expected background has an uncertainty, {\sc Adroit} allows its normalisation to vary during the fit, while requiring it to remain positive. Specifically, we take
\beq
 \ln C(\theta)=-\frac{(\theta-1)^2}{2\sigma_\theta^2},
 \qquad
 \sigma_\theta=
 \begin{cases}
  \sigma_-, & \theta\leq1,\\
  \sigma_+, & \theta>1,
 \end{cases}
\eeq
Here $\sigma_-$ and $\sigma_+$ are the lower and upper uncertainties on the background count, expressed as fractions of $B$.
When no uncertainty is given, the background normalisation is fixed to $\theta=1$.

For a specified spectrum, we assess whether adding a signal improves the description of the data by comparing the best fit with signal and background to the best fit with background alone. This comparison is quantified via the profile likelihood-ratio test statistic \cite{Cowan:2010js}
\beq
 q_0=2\Big[\ln\mathcal L(\hat s,\hat\theta)
       -\ln\mathcal L(0,\hat\theta_0)\Big],
 \label{eq:q0}
\eeq
with $\hat s\geq0$. The hatted parameters indicate the best-fit values, i.e.~those that maximise the likelihood.
Here $\hat s$ and $\hat\theta$ give the best fit when a signal is included, while $\hat\theta_0$ gives the best background-only fit. A larger $q_0$ indicates a stronger preference for adding a signal. This is a local test since it concerns one specified signal hypothesis, including its mass.
For the signal alone, the best-fitting candidate is the one with the largest $q_0$. 

Searching many candidate spectra increases the chance that a background fluctuation will resemble a signal. To account for the look-elsewhere effect \cite{Gross:2010qma}, \textsc{Adroit} generates simulated datasets containing only background events.
For a fixed background normalisation, the number of events is drawn from a Poisson distribution with mean $B$, and their energies from $b(E)/B$. When the background normalisation is uncertain, the simulations use its background-only best-fit value and repeat the treatment of its uncertainty.
For each simulated dataset, we repeat the candidate search and record the largest $q_0$. We then determine how often this reaches or exceeds the largest value found in the observed data.
The resulting global significance applies to the specified search procedure and background model.
This procedure for determining the global significance is analogous to that employed in dark matter direct detection searches, such as the recent LZ analysis \cite{LZ:2026axp}.

Several interactions can describe a small recoil sample almost equally well. To estimate how many signal events would help distinguish them, we compare their predicted energy distributions. For this comparison, we use efficiency-weighted recoil spectra before energy-resolution smearing. For two such shapes $f$ and $g$, normalised over the recoil-energy window $W_R$, we use the Jeffreys divergence:
\beq
 J(f,g)=\int_{W_R}(f-g)\ln(f/g)\,\dd E_R.
\eeq
A small value of $J$ indicates similar spectra, for which a larger event sample would be needed to distinguish the two predictions.\footnote{In evaluating the logarithm numerically, we replace values below $10^{-12}$ of each spectrum's maximum by that threshold. This keeps the comparison finite when one spectrum vanishes over part of the energy range.}

Specifically, in line with \cite{Kullback}, we take $J$ as a measure of how distinguishable the two spectra are, and adopt $N_{2\sigma}=4/J$ as a heuristic estimate of the number of signal events needed to separate them at approximately $2\sigma$.
This estimate uses the signal shapes alone and omits backgrounds, energy-resolution smearing and uncertainties in the inputs.
For a given number of signal events $N$, we group candidate interactions whose predicted spectra would be difficult to distinguish into spectral classes, using $N\leq N_{2\sigma}$ to identify similar pairs. Two candidates can belong to the same group through a chain of similarities, even if their own spectra are distinguishable. We test this procedure using simulated recoil samples generated from known interactions, as described in Appendix \ref{A}. For these samples, we report separately whether the highest ranked candidate is the generating interaction, has a similar spectrum, or belongs to the same spectral class.

Candidates from the leading spectral classes are taken forward to the final ranking. We rank these candidates, together with the background-only hypothesis, using the Akaike information criterion \cite{Akaike1974}
\beq
\mathrm{AIC}=2k-2\ln\mathcal L_{\max},
\eeq
where $k$ is the number of fitted parameters. Here each fixed grid point is treated as a separate candidate, and $k$ counts the parameters fitted within that candidate. This penalises additional parameters unless they sufficiently improve the fit.
Candidates are ranked by their AIC value, with smaller values preferred. When a signal candidate ranks first, we also retain candidates with $\Delta\mathrm{AIC}\leq2$, measured relative to that minimum. If the background-only hypothesis ranks first, the analysis reports that outcome. 

When null searches are not included, ranking candidates with the same number of fitted parameters by $q_0$ or AIC gives the same ordering. 
When null searches are included, candidates are ranked using a joint fit with a common interaction strength, incorporating these searches via likelihoods reconstructed from their upper limits. The AIC then uses the maximum joint likelihood. A candidate that fits the signal well can therefore fall in the ranking if the required strength conflicts with a null result. Without null searches, the ranking uses the signal alone. We also calculate a parameter-goodness-of-fit statistic, whose $p$-value is calculated from a one-degree-of-freedom chi-squared approximation \cite{Maltoni:2003cu}. This measures tension between datasets and is distinct from the global significance of the excess. This $p$-value is reported alongside the ranking, rather than used as an additional ranking score. Relic density and other complementary constraints are also reported separately from the ranking.

The second path ``explain a signal'' starts from a specified signal and particle physics model, with the latter being defined by the researcher (manually or via an agent) or selected from the model library. 
At each dark matter mass, \textsc{Adroit} fits the model's physical coupling to the events. The coupling fixes both the expected number of signal events and the shape of their energy spectrum, so the signal is not given a free normalisation. The fit maximises the event likelihood, while the model's other parameters are held fixed or are determined by specified relations between the model's masses and couplings.
Where possible, \textsc{Adroit} first fits only the number of signal events, keeping the spectral shape fixed, and converts that number into the coupling that would produce it. The full prediction at that coupling is then checked against the fit. If the check fails, for example because the coupling also changes the spectral shape, or if the user has specified a coupling range, the coupling is fitted directly within its allowed range, recalculating the signal at each step.
Repeating the fit across masses gives a curve in the mass-coupling plane. A mass is kept on the curve only if the model describes the events better than background alone.
{\sc Adroit} subsequently calculates the relic abundance and evaluates the complementary constraints at the fitted points. Where the data admit several interpretations, the analysis retains these alternatives for further investigation.


\section{Worked Examples}\label{S5}

In this section we illustrate the different ways of using \textsc{Adroit} through three worked examples (see Table~\ref{tab:workflows}). These show how the system can
help construct a particle model, test it against experimental constraints, and explore possible explanations of a dark matter signal. For all language-model calls made to produce these examples we used Claude Opus~5.5.

\begin{table}[b]
\centering
\begin{tabular}{>{\raggedright\arraybackslash}p{.21\linewidth}>{\raggedright\arraybackslash}p{.28\linewidth}>{\raggedright\arraybackslash}p{.43\linewidth}}
\toprule
Task & Input & Main output \\
\midrule
Signal best fits & LZ extended-window event & Ranked recoil hypotheses and comparisons with selected null searches. \\[8pt]
Write a model & Text description of a Stueckelberg $B-L$ model & Fields, interactions and explicit model assumptions (exportable as UFO). \\[8pt]
Constrain a model & Generated $B-L$ model & Parameter plane with relic density curve and experimental constraints. \\[8pt]
Explain a signal & $Z'$ portal and LZ event & Parameter plane with signal fits, relic density curve, and constraints. \\
\bottomrule
\end{tabular}
\vspace{-4mm}
\caption{The four tasks illustrated by the three examples. \label{tab:workflows}}
\end{table}

\subsection{Signal best fits: the LZ extended window}\label{sec:lz-example}

As an example, we ask which interactions are consistent with the single high-energy nuclear-recoil-like event observed in the LZ extended-window search~\cite{LZ:2026axp}. Its reconstructed energy is $248\pm23\,\mathrm{(stat)}\pm23\,\mathrm{(sys)}$ keV, we combine the two uncertainties into a Gaussian width of 32.5 keV. The background density of our likelihood is fixed to reproduce the published $3.4\sigma$ local significance of the isoscalar $L_{10}$ interaction \cite{Anand:2013yka} at 1~TeV. This provides an approximate recoil-energy likelihood in place of the collaboration's full two-dimensional event fit. It includes the high-energy fall in detection efficiency (by digitising the curve in Figure S2 of \cite{LZ:2026axp}), setting the efficiency to zero beyond its stated range near 300 keV.

The scan fits 612 hypotheses, including the non-relativistic operators $\mathcal O_1$ and $\mathcal O_4$, elastic relativistic-interaction benchmarks and inelastic splittings. We select CRESST-III, CDMSlite, PICO-60 and DEAP-3600~\cite{CRESST:2019jnq,SuperCDMS:2017nns,PICO:2019vsc,DEAP:2019yzn} 
as complementary null searches. As in the experimental analyses, the CRESST-III and CDMSlite limits use the optimum interval method \cite{Yellin:2002xd}. The run also generates 200,000 background-only simulations using this approximate likelihood to obtain a measure of the global significance under the  background-only assumption.
 The run can be manually implemented in {\sc Adroit} via the terminal command 
\begin{quote}\footnotesize
\begin{verbatim}
Adroit signal --experiment 'LZ extended window (2026)' \
  --toys 200000 --profile none \
  --null 'CRESST-III 2019,CDMSlite Run 2,PICO-60 C3F8 (2019),DEAP-3600 (2019)' \
  --out lz_example
\end{verbatim}
\vspace{-1mm}
\end{quote}
Here \texttt{-{}-profile none} runs the code without language-model calls. The same study can be configured via the graphical interface, with agent assistance for the inputs. The output lists the fitted interactions in order of AIC from the joint signal and null-search fit, together with their spectral classes and the available null-search comparisons. To form these groups, we relate interactions whose predicted recoil spectra would be difficult to distinguish in a hypothetical sample of 20 signal events (the actual fit uses the single observed LZ event).

Table~\ref{tab:lz-ranking} shows the highest ranked entries. The top-ranked entry is isovector $\mathcal O_1$, with $m_\chi=1000$ GeV and splitting $\delta=350$ keV, but the small score differences among the four do not establish a preference for one interaction over the others.
Notably, these candidates are sensitive to the high-speed tail of the halo distribution, and thus their inferred couplings are particularly sensitive to the halo assumptions. Because the interactions are inelastic, scattering becomes impossible for sufficiently low maximum speeds. Notably, the preferred splitting, 350 keV, is the largest value tested.

The selected null searches do not constrain these candidates. For the adopted halo assumptions, their 350 keV inelastic splitting prevents scattering on the lighter nuclei, while the tungsten recoils lie above the CRESST-III analysis window. Thus, including the selected null searches does not change the fit scores of these four candidates.

Finally, the background-only simulations give a global significance under the approximate likelihood. However, its background is calibrated to one published local significance, rather than reconstructed from the collaboration's background model, and it does not reproduce the collaboration's global significance. We therefore do not quote this value and retain LZ's reported $2.6\sigma$~\cite{LZ:2026axp}.

\begin{table}[t]
\centering
\begin{tabular}{cccccc}
\toprule
Interaction & Isospin & $m_\chi$ (GeV) & $Z_{\rm local}$ ({\sc Adroit}) & $Z_{\rm local}$ (LZ) & ~~~~$\Delta\mathrm{AIC}$ \\
\midrule
$\mathcal O_1$ & Isovector & 1000 & 3.65 & 3.4 & 0     \\
$\mathcal O_1$ & Isovector & 4000 & 3.64 & 3.4 & 0.055 \\
$\mathcal O_4$ & Isovector & 1000 & 3.62 & 3.4 & 0.195 \\
$\mathcal O_4$ & Isoscalar & 1000 & 3.62 & 3.4 & 0.196 \\
\bottomrule
\end{tabular}
\vspace{1mm}\caption{
The four leading recoil fits identified by \textsc{Adroit} for the one-event approximation to the LZ analysis. All have an inelastic splitting $\delta=350$ keV. For \textsc{Adroit}, $Z_{\rm local}=\sqrt{q_0}$ is the asymptotic local significance, with $q_0$ defined in Eq.~(\ref{eq:q0}). This is compared with the simulation-calibrated value obtained by LZ using the full likelihood (Table~S8 of \cite{LZ:2026axp}). The $\Delta\mathrm{AIC}$ values correspond to \textsc{Adroit}'s approximate likelihood.
 \label{tab:lz-ranking}}
\end{table}

\subsection{Build and constrain a model: Stueckelberg $B-L$}\label{sec:model-building-example}

We next follow a model from its construction to its comparison with data. We give {\sc Adroit}'s model-building agent the following prompt:
\begin{quote}
\textit{A gauged $U(1)_{B-L}$ with a Stueckelberg $Z'$ at 5 TeV, coupling around $g = 0.2$. With Dirac DM with $B-L$ charge 1, mass around 2 TeV.}
\end{quote}

\begin{figure}[h]
\centering
\includegraphics[width=.82\linewidth]{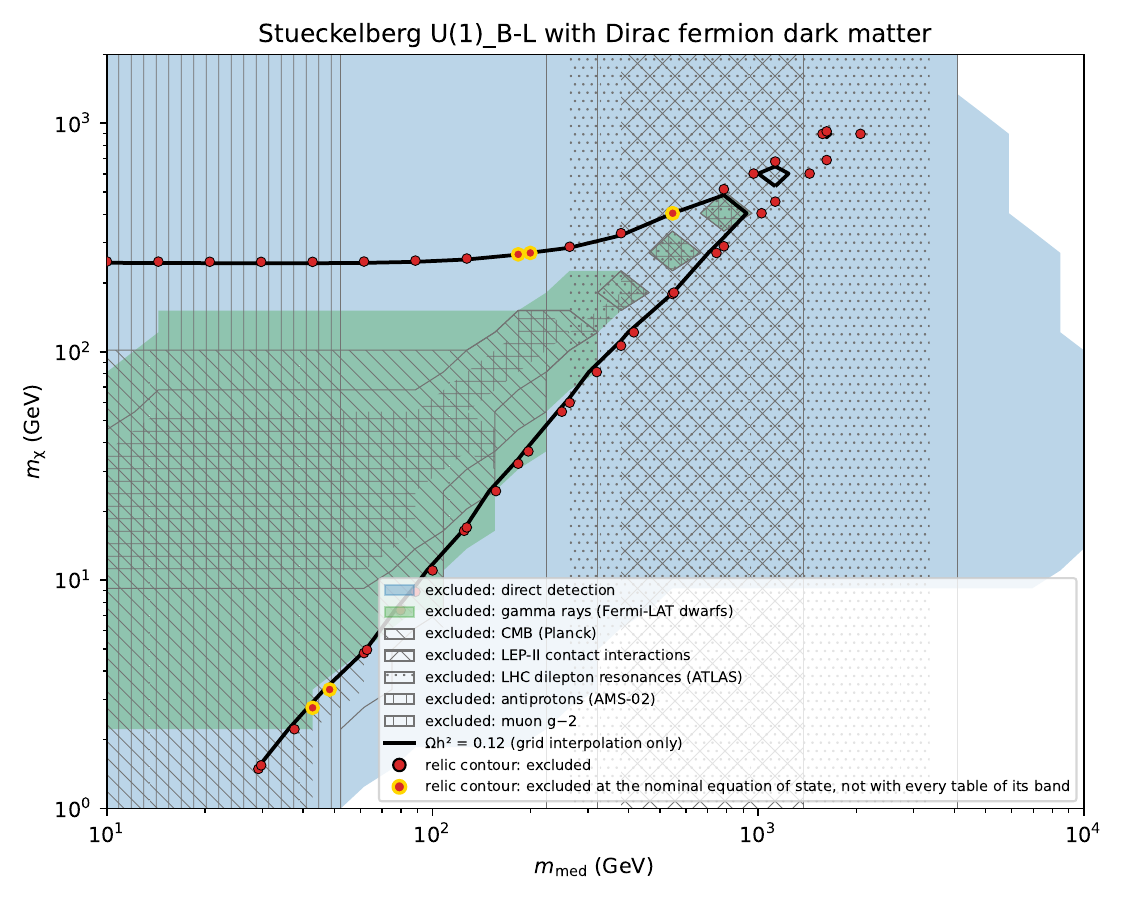}
\vspace{-4mm}
\caption{Stueckelberg $B-L$ parameter plane for $g_{B-L}=0.2$, as output by {\sc Adroit}. Experimental exclusion are displayed: shading for direct detection and the Fermi-LAT dwarf galaxies, and hatching for the CMB~\cite{Planck:2018vyg}, the LEP-II contact-interaction bound \cite{ALEPH:2013dgf}, the ATLAS dilepton search~\cite{ATLAS:2019erb}, AMS-02 antiprotons \cite{Calore:2022stf} and the muon $g-2$ \cite{Muong-2:2025xyk}. The black line is the $\Omega_\chi h^2=0.12$ contour interpolated on the $20\times20$ grid, and circles mark the 48 individually refined thermal solutions, all of which are excluded. Gold outlines identify six solutions with unresolved additional checks using alternative equation-of-state tables (they do not indicate surviving solutions). The approximations behind each solution's verdict are listed in {\sc Adroit}'s report. No relic abundance is computed at $m_\chi=1$ GeV, where low-energy hadronic annihilation is outside the calculation's supported range.}
\label{fig:bl-plane}
\vspace{-3mm}
\end{figure}

The builder returns a neutral vector mediator, a stable Dirac fermion $\chi$ and three right-handed neutrinos, together with the $B-L$ couplings of the Standard Model fermions. The researcher approves this model for the calculation. The relevant Lagrangian terms are
\beq
 \mathcal L &\supset -\frac14 Z'_{\mu\nu}Z'^{\mu\nu}
 +\frac12 m_{Z'}^2 Z'_\mu Z'^{\mu}
 +\bar\chi(i\slashed{\partial}-m_\chi)\chi
 +g_{B-L}Z'_\mu\bigl(\bar\chi\gamma^\mu\chi+J^\mu_{B-L}\bigr),
 \label{eq:bl-model}
 \eeq
where $Z'_{\mu\nu}$ is the mediator field strength with
 \beq
 J^\mu_{B-L}&=\sum_{i=1}^{3}\left[
 \frac13\left(\bar Q_i\gamma^\mu Q_i+\bar u_{Ri}\gamma^\mu u_{Ri}
 +\bar d_{Ri}\gamma^\mu d_{Ri}\right)
 -\bar L_i\gamma^\mu L_i-\bar e_{Ri}\gamma^\mu e_{Ri}
 -\bar\nu_{Ri}\gamma^\mu\nu_{Ri}\right],
 \label{eq:bl-current}
\eeq
and where $Q_i$ and $L_i$ are the left-handed quark and lepton doublets. The mediator mass is a Stueckelberg input, with no additional Higgs scalar. The generated model assigns the right-handed neutrinos $B-L=-1$ and takes the Dirac neutrino masses to be 0.1 eV, sets kinetic mixing with hypercharge to zero, and imposes a $Z_2$ symmetry under which only $\chi$ is odd. These are assumptions made by the builder in addition to the prompt, they are displayed with the generated model for the researcher to review and approve (or modify).

We then take the approved model into the \emph{Constrain a model} workflow. The prompt specifies a benchmark at $m_{Z'}=5$ TeV and $m_\chi=2$ TeV. For the parameter-space study, the system varies $10\leq m_{Z'}/{\rm GeV}\leq10^4$ and $1\leq m_\chi/{\rm GeV}\leq2000$ on a logarithmic $20\times20$ grid, holding $g_{B-L}=0.2$ fixed. 
The system assumes standard thermal freeze-out and uses the SHM uncertainty region for direct detection. The same interaction governs annihilation into fermions and coherent nuclear scattering, so a coupling that gives the required abundance can also produce a strong direct detection signal.
Accordingly, the scan compares each point with direct-detection limits from LZ~\cite{LZ:2024zvo}, Fermi-LAT observations of dwarf galaxies, and LHC searches for high-mass dilepton resonances~\cite{ATLAS:2019erb}. It also evaluates the CMB \cite{Planck:2018vyg}, cosmic-ray antiprotons from AMS-02 \cite{Calore:2022stf}, 
the muon $g-2$  \cite{Muong-2:2025xyk}, solar capture \cite{IceCube:2016dgk},  gamma-ray lines, the LEP-II contact-interaction bound~\cite{ALEPH:2013dgf}, LHC dijet and recast searches, the muon $g-2$ and perturbativity.
The gamma-ray comparison uses the published six-year Fermi-LAT limits~\cite{Fermi-LAT:2015att}. In cases for which the annihilation final states do not match a published limit, it is supplemented by a spectral recast of the 14-year dwarf-galaxy data~\cite{McDaniel:2023bju}.  The six-year Fermi-LAT limits exclude 75 sampled points \cite{Fermi-LAT:2015att},
and the analysis of the 14-year data excludes a further 12 \cite{McDaniel:2023bju}.
The gamma-ray comparison does not exclude 222 points and remains unresolved at the remaining 91.

For the generated model, a scan is initiated (without needing a live call) via the command
\begin{quote}\small
\begin{verbatim}
Adroit track1 examples/bl_model.yaml \
  --x m_med:10:10000:log:20 --y m_chi:1:2000:log:20 \
  --halo preset --propagation conservative --profile none \
  --out bl_example
\end{verbatim}
\end{quote}
Here \texttt{track1} denotes the model-to-constraints workflow and \texttt{-{}-propagation conservative} selects the weakest charged-cosmic-ray bounds obtained from the adopted variations in Galactic transport and dark matter density profiles.  The calculation identifies 48 points that reproduce the observed relic abundance via freeze-out, however, all 48 thermal solutions are excluded by direct detection for the central halo assumptions. 
 This exclusion is also established across the sampled halo region for 45 points. For the remaining three, some halo-dependent direct-detection comparisons are incomplete, but their exclusion is independently established by the ATLAS dilepton comparison \cite{ATLAS:2019erb}. Thus all resolved relic density solutions in the scanned plane are excluded for $g_{B-L}=0.2$. The {\sc Adroit} output summarising this information is presented in Figure \ref{fig:bl-plane}.

\vspace{1mm}
\subsection{Explain a signal: Axial $Z'$ portal}
\vspace{1mm}

We finally ask whether a specific particle model can explain the LZ recoil event. For our example we consider Dirac dark matter $\chi$ interacting with quarks through a new axial-vector mediator $Z'$ \cite{Lebedev:2014bba}. The relevant interactions are
\beq
 \mathcal L_{\rm int}
 = Z'_\mu\Big(g_\chi\bar\chi\gamma^\mu\gamma^5\chi
 +g_q\sum_q\bar q\gamma^\mu\gamma^5q\Big).
 \label{eq:axial-model}
\eeq
We fix $m_{Z'}=1$ TeV and $g_q=0.25$, the lepton couplings are set to zero. This is a simplified model, with no mediator mixing or additional states responsible for its mass included in the calculation (we note in passing that UV completion of such mediators is rather unsatisfactory \cite{Ismail:2016tod}).
We pass this model to the \emph{Explain a signal} workflow and fit $g_\chi$ within $0\leq g_\chi\leq1.5$ at sixteen logarithmically spaced dark matter masses between 5 and 1000 GeV. The LZ fit uses the SHM with the halo parameters adopted by the experiment. For halo-dependent constraints, we explore the SHM uncertainty region described in Section \ref{S3}. We assume standard freeze-out and take $\chi$ to constitute all of the dark matter.

\begin{table}[b]
\centering
\begin{tabular}{cccc}
\toprule
~~~$m_\chi$ (GeV)~~~ & ~~~$g_\chi$ ~~~& ~~~$\Omega_\chi h^2$~~~ & ~~~$\sqrt{q_0}$~~~ \\
\midrule
393.2 & 0.608 & 0.120 & 2.61 \\[8pt]
626.9 & 0.748 & 0.120 & 2.69 \\
\bottomrule
\end{tabular}
\vspace{2mm}
\caption{Points on the axial $Z'$ signal-fitted curve that reproduce the observed thermal relic abundance. The local fit statistic is expressed as $\sqrt{q_0}$, using the asymptotic conversion of Eq.~(\ref{eq:q0}). Both points are excluded by the CMS dijet search~\cite{CMS:2018mgb}. \label{tab:axial-thermal}}
\vspace{2mm}
\end{table}

\begin{figure}[t]
\centering
\includegraphics[width=.8\linewidth]{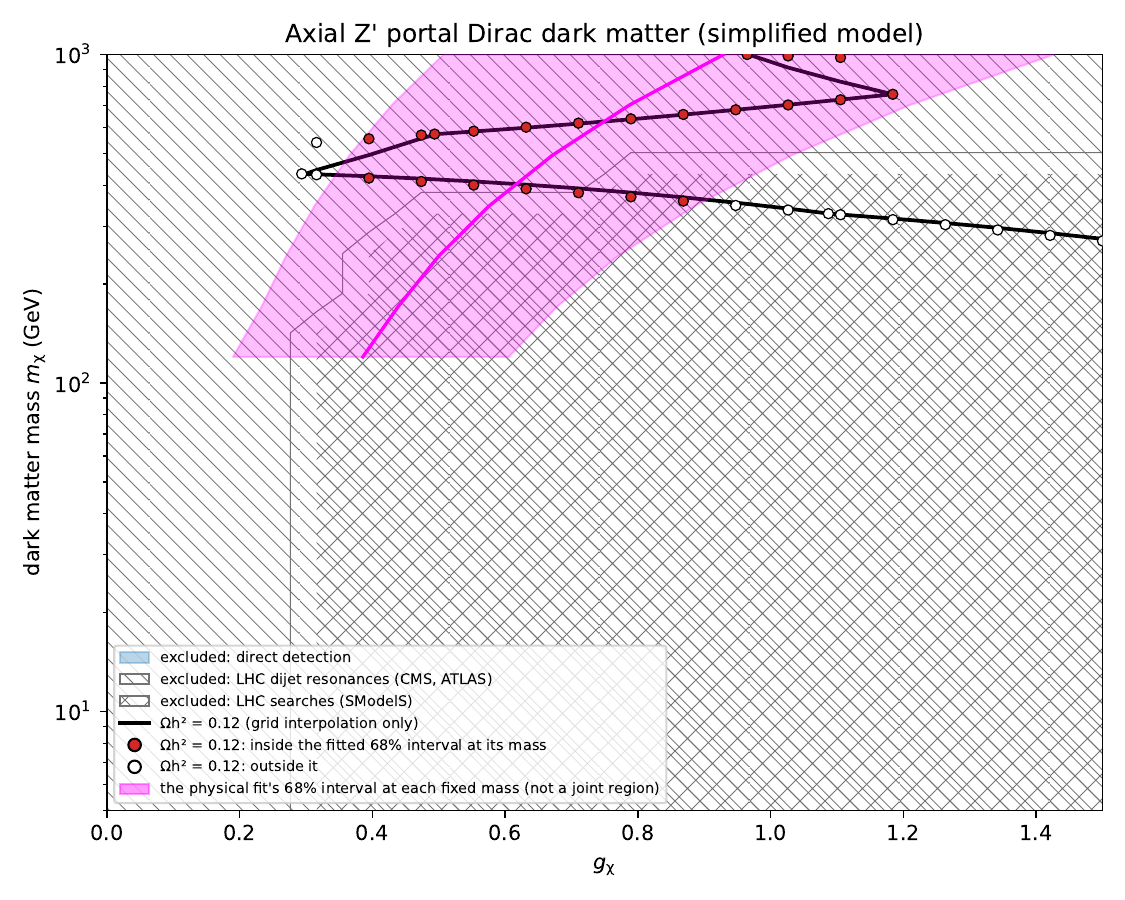}
\vspace{-3mm}\caption{Axial $Z'$ model at $m_{Z'}=1$ TeV and $g_q=0.25$, as output by {\sc Adroit}. The magenta curve shows the best-fitting coupling at each dark
matter mass, and the shaded band gives its 68\% likelihood interval. The scan covers 5-1000 GeV, but only fits with $q_0\geq1$ are plotted. The first sampled mass satisfying this criterion is $m_\chi\simeq120$ GeV. The black line is the $\Omega_\chi h^2=0.12$ contour interpolated on the grid, and circles are the 34 refined thermal solutions: red (22) where they lie inside the 68\% interval at their mass, white (12) outside it. Hatching shows the experimental exclusions. The CMS dijet search~\cite{CMS:2018mgb} excludes the entire plane, including all 34 thermal solutions and both intersections of Table~\ref{tab:axial-thermal}. The LHC searches recast by SModelS exclude the lower-right region. Direct detection excludes only one sampled point, at $(m_\chi,g_\chi)\simeq(35~{\rm GeV},1.5)$, too small to appear as shading.
\label{fig:axial-plane}\vspace{-2mm}}
\end{figure}

The coupling fit at each mass gives the magenta curve in Figure~\ref{fig:axial-plane}, its shaded band
 shows the 68\% likelihood interval at each fixed mass. The broad band reflects the limited information in a single event.
The best fit within the sampled range occurs at its upper boundary, for $m_\chi=1000$ GeV and $g_\chi=0.929$, with numerical fit score $\sqrt{q_0}=2.73$. 

{\sc Adroit} then calculates the relic abundance along the fitted curve. Annihilation becomes more efficient near the mediator resonance, $2m_\chi=m_{Z'}$, which can be seen in Figure \ref{fig:axial-plane}. Refining the mass and repeating the signal fit identifies two points that reproduce $\Omega_\chi h^2=0.12$, at $(m_\chi,g_\chi)\simeq(393~{\rm GeV},0.608)$ and $(627~{\rm GeV},0.748)$. Specifically at these points the signal-fitted curve and the relic density curve intersect. Both predict approximately one signal event in the LZ window. Their parameters and fit statistics are given in Table~\ref{tab:axial-thermal}.

The scan identifies 34 thermal relic solutions, shown as circles in Figure \ref{fig:axial-plane}. 
Notably, obtaining the observed abundance and fitting the recoil event are both possible in this model. 
However, the same quark coupling that provides the LZ recoil also allows the $Z'$ to be produced at the LHC and decay into two jets. The CMS search for dijet resonances \cite{CMS:2018mgb} excludes all sampled points for this choice of mediator mass and quark coupling, including both thermal points on the signal-fitted curve. The dark matter scattering cross section is spin dependent, and the direct-detection searches included here do not exclude any of the 34 thermal solutions. They exclude only one sampled grid point, so no shaded direct-detection exclusion region is visible. Although the model can reproduce the observed abundance while improving the fit to the recoil event over background alone, the thermal solutions found here are excluded by the experimental constraints.


\section{Concluding Remarks}
\label{S6}

In this work we have presented a new tool for dark matter model building and comparison to data:  \textsc{Adroit}.
The system is designed to aid experienced physicists by leveraging AI in the development and testing of dark matter models.
Specifically, it connects model construction and literature searches to recoil fits, relic abundance calculations, and complementary experimental constraints.
Established numerical tools evaluate the physical predictions, while language-model agents assist with preparing the inputs and interpreting the results.
 A researcher can begin with a library model, a new Lagrangian, or an event sample, and follow the calculation through to the resulting constraints and report, with an optional AI-assisted summary.
The worked examples illustrate the workflow via which an idea is formulated into a concrete Lagrangian and then tested against constraints and signals.
Planned extensions include additional library models and broader coverage of experimental constraints. 

\vspace{4mm}
\noindent {\bf Data and software availability.}
Available at: \url{https://github.com/jamesunwin/adroit} 

\vspace{4mm}
\noindent {\bf Acknowledgments.}
The author is grateful to Oxford's Rudolf Peierls Centre, New College, and Queen's College for their hospitality. OpenAI Codex and Anthropic Claude assisted in the software development.

\appendix

\section{Validation}
\label{A}

In this Appendix we show comparisons between the calculations from {\sc Adroit} and independent numerical codes and published experimental results. We also use simulated recoil samples to test how well the analysis identifies known interactions.

\begin{table}[b]
\centering
\begin{tabular}{cccccc}
\toprule
Benchmark & ~$m_{H^0}$ [GeV]~ & ~$m_{A^0}$ [GeV]~  & ~$m_{H^\pm}$ [GeV]~  & ~$\lambda_{345}$~ & $\Omega_{\rm Adroit}/\Omega_{\rm micrOMEGAs}$ \\
\midrule
BM5 & 100 & 105 & 200 & 1.000 & 0.986 \\[4pt]
BM6 & 100 & 105 & 200 & 0.002 & 0.993 \\[4pt]
HP2 & 557 & 562.316 & 565.417 & $-0.1385$ & 0.980 \\[4pt]
HP8 & 933 & 939.968 & 943.787 & $-0.2435$ & 0.982 \\[4pt]
HP10 & 990 & 992.360 & 998.120 & $-0.0510$ & 0.984 \\
\bottomrule
\end{tabular}
\caption{Relic density comparison with micrOMEGAs at five inert-doublet benchmarks of  \cite{Belyaev:2016lok,Kalinowski:2018ylg}. The inert-doublet dark matter is $H^0$ and $\lambda_{345}$ is the portal coupling. The two calculations use the same electroweak inputs, $m_h=125.09$ GeV and $\lambda_2=0.5$. \label{TabA1}}
\end{table}

\subsection{Relic-abundance calculations}

We first compare the relic abundance predicted by {\sc Adroit} with micrOMEGAs 7.1.10. Specifically, we study five inert-doublet benchmarks \cite{Belyaev:2016lok,Kalinowski:2018ylg} with matched masses, couplings and electroweak inputs. The results agree to within $\sim2$\%, as shown in Table \ref{TabA1}. These benchmarks were used during development, including to investigate discrepancies arising from different input choices.
We also compare the constant $s$-wave annihilation cross-section \(\langle\sigma v\rangle\) required to obtain $\Omega_\chi h^2=0.12$ with the Majorana dark matter model of \cite{Saikawa:2020swg}. The required cross-sections agree to within 0.47\% over the tested dark matter mass range of 1-1000~GeV.

\subsection{Comparisons with experimental results}

We check whether \textsc{Adroit} reproduces the main results of published experimental analyses. For exclusion limits, we ask whether the code rules out the same interaction strengths as the experimental analysis at a given dark matter mass. For signal fits, we ascertain whether it finds the same preference for a dark matter signal over background alone.
We compare \textsc{Adroit}'s exclusion limits with the published limits at selected dark matter masses. 
We take a calculated limit as in agreement with the published one if the two differ by no more than a factor of 1.3, or if they agree once the mass is shifted by up to 3\%, which allows for the steep rise of the limits near experimental thresholds. We first used this test to identify the mass range over which each experiment's limit is reliably reproduced. In a later check at 81 masses, all 69 points inside those ranges agreed with the published limits, as did two of the twelve points outside them.

For the LZ  high-recoil event \cite{LZ:2026axp}, we use the published extended window results to test
how the interpretation of the recoil event changes with the
assumed dark matter interaction. The tables in \cite{LZ:2026axp} give local
significances for 612 combinations of interaction, mass and
inelastic splitting. Each significance measures how strongly
that particular hypothesis improves the fit over background
alone. 
We fix the approximate background density using a single published benchmark, and then test the result on 304 of the remaining benchmarks. For 95.1\% of these, the calculated significance lies within $0.3\sigma$ of the published value. However, 163 of these have a published significance of exactly zero, because the model predicts essentially no signal at the observed energy, and all of these are trivially reproduced. Among the 141 remaining benchmarks, 89.4\% agree to within $0.3\sigma$.
Thus, the simplified treatment
reproduces the published local significances across most
of the tested hypotheses.

\subsection{Recovery of interactions from simulated signals}

We validate the system on simulated signal datasets, whose event energies are drawn from the recoil spectra predicted by Eq.~(\ref{eq:rate}). Specifically, we construct 384 datasets by selecting a candidate from the interaction library and drawing 3, 5, 10, 20 or 50 signal events from its predicted recoil-energy spectrum, including detector
efficiency. We also add background events with a Poisson-distributed count. We consider xenon, calcium tungstate, germanium and argon targets. 
For this test, we fit all signal candidates to each simulated dataset and retain those with  $\Delta{\rm AIC}\leq 2$, relative to the best-fitting signal candidate (irrespective of where the background-only hypothesis would rank).
Because the interaction used to generate each sample is known, we can assess what the fit recovers.

\begin{figure}[t]
\centering
\includegraphics[width=.78\linewidth]{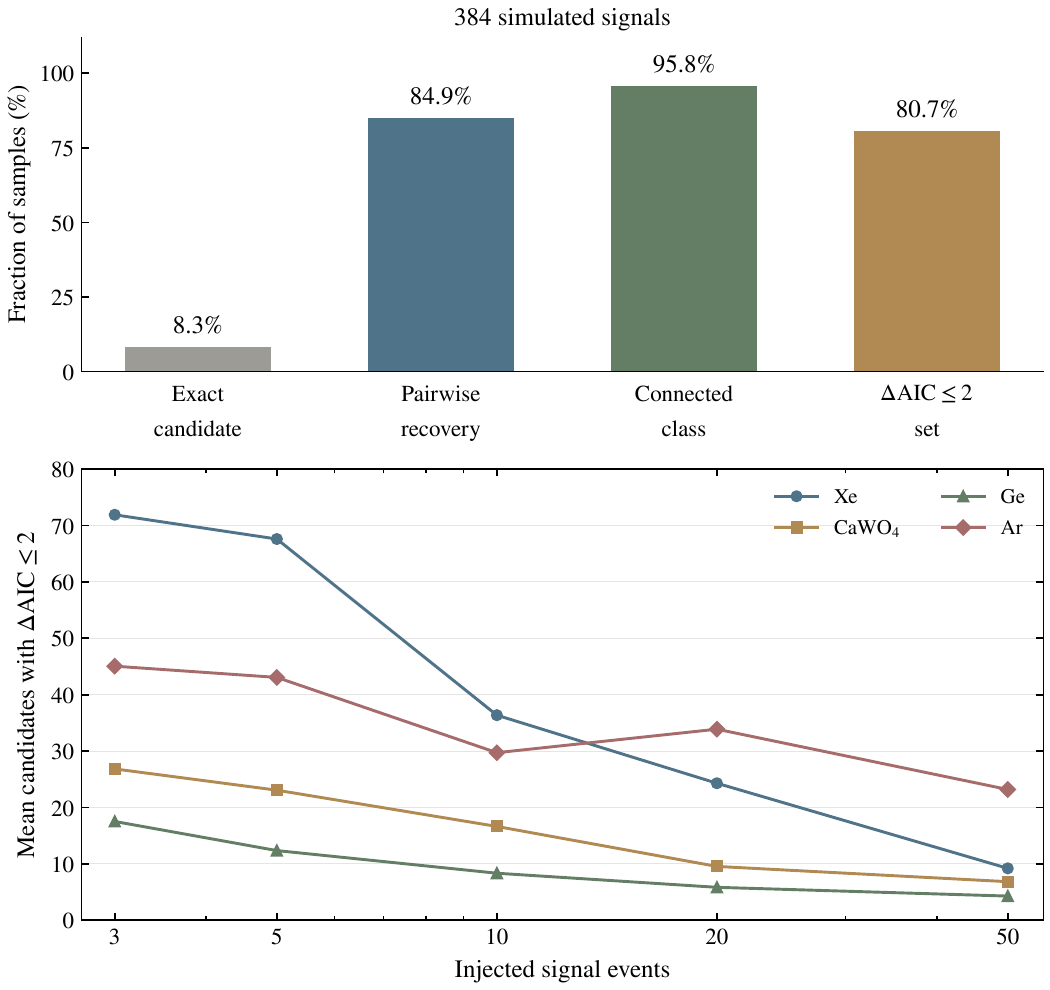}
\vspace{-3mm}
\caption{
From 384 simulated recoil samples based on a given dark matter interaction, the plots indicate how successful {\sc Adroit} is at correctly identifying the injected signal.
{\bf Upper:} The fractions of samples for which the generating interaction i).~matched the highest ranked candidate, ii).~was pairwise similar to the highest ranked candidate, iii).~belonged to the same connected spectral class as the highest ranked candidate, or iv).~was within the $\Delta\mathrm{AIC}\leq2$ set. {\bf Lower:} The mean number of candidates in the $\Delta\mathrm{AIC}\leq2$ set as a function of the injected signal count for different materials. The candidate libraries contain 612 (xenon), 196 (calcium tungstate), 140 (germanium) and 168 (argon) mass-interaction hypotheses. Hypotheses that predict no signal in a target are excluded from its fitted set, leaving 612, 196, 40 and 48 respectively. }
\label{fig:injections}
\vspace{-2mm}
\end{figure}

For this study, we form the spectral classes using each sample's injected signal count. We record whether the highest ranked candidate is the generating interaction, whether their spectra satisfy the pairwise similarity criterion, and whether they belong to the same spectral class.
Figure \ref{fig:injections} illustrates the results of these tests, showing rates and the mean number of retained candidates for each target and signal count. 
Across the 384 samples, the highest ranked candidate matches the generating interaction in 8.3\% of cases.  It satisfies the pairwise similarity criterion in 84.9\% of cases and belongs to the same spectral class in 95.8\%. This latter value requires some care, because with few events the classes are very broad. With three events nearly all xenon candidates fall into a single class, and for argon this is true at every signal count. As a result, a candidate picked at random would land in the correct class 68.7\% of the time. The class result is therefore better read as showing that related explanations are retained than as identifying the interaction. 

Moreover, it can be seen that the advantage over a random pick emerges as the sample grows. With 50 events the correct class is recovered in 90\% of xenon samples, against 31\% at random, and in all germanium samples, against 14\%. The $\Delta\mathrm{AIC}\leq2$ set contains the generating interaction in 80.7\% of cases. Notably, the number of competing explanations generally decreases as the signal count increases. Nevertheless, identifying the exact interaction remains difficult. These results illustrate the value of retaining several plausible explanations for further investigation. Measurements with different target nuclei and complementary experimental constraints can then help further distinguish them.

\subsection{Comparison with direct language-model responses}

Finally, we compare the results of {\sc Adroit} analyses with those obtained by asking a language model to interpret the signal events directly. We use simulated
recoil samples, for which the interaction that generated
the events is known (following the procedure above). The direct prompts give the event
energies and candidate labels, but no calculated recoil
spectra. 

We assess whether each method's preferred candidate
belongs to the same spectral class as the generating
interaction: it need not identify the exact interaction,
but should recover one in the same spectral class.
On the 56 cases completed by both language models, the {\sc Adroit} workflow recovers this class in 94.6\% of cases, compared with 80.4\% for direct prompting of Claude Opus 5.5 and 75.0\% for GPT-5.6 Sol. For the stricter pairwise criterion the rates are 82.1\%, 50.0\% and 39.3\%. The {\sc Adroit} result is a deterministic fit and does not depend on the language model.

\end{document}